\documentclass[conference]{IEEEtran}

\ifCLASSINFOpdf
\else
\fi
\usepackage[cmex10]{amsmath}

\usepackage{graphicx,epic,eepic,epsfig,amsmath,latexsym,amssymb,mathrsfs,verbatim,subfigure,color}

\newcommand{\bra}[1]{\langle#1|}
\newcommand{\ket}[1]{|#1\rangle}

\def\squareforqed{\hbox{\rlap{$\sqcap$}$\sqcup$}}
\def\qed{\ifmmode\squareforqed\else{\unskip\nobreak\hfil
\penalty50\hskip1em\null\nobreak\hfil\squareforqed
\parfillskip=0pt\finalhyphendemerits=0\endgraf}\fi}
\def\endenv{\ifmmode\;\else{\unskip\nobreak\hfil
\penalty50\hskip1em\null\nobreak\hfil\;
\parfillskip=0pt\finalhyphendemerits=0\endgraf}\fi}

\usepackage{cite}

\long\def\ignore#1{}

\def\C{\mathbb{C}}

\begin{document}
%
\title{Codeword Stabilized Quantum Codes for Asymmetric Channels}

\author{\IEEEauthorblockN{Tyler Jackson\IEEEauthorrefmark{1}\IEEEauthorrefmark{2}, 
Markus Grassl\IEEEauthorrefmark{3}\IEEEauthorrefmark{4}, and 
Bei Zeng\IEEEauthorrefmark{1}\IEEEauthorrefmark{2}\IEEEauthorrefmark{5}}\medskip
\IEEEauthorblockA{
\IEEEauthorrefmark{1}Department of Mathematics $\&$ Statistics,
University of Guelph, Guelph, Ontario, N1G 2W1, Canada\\
\IEEEauthorrefmark{2}Institute for Quantum Computing, University of Waterloo,
Waterloo,  Ontario, N2L 3G1, Canada\\
\IEEEauthorrefmark{3}Institut f\"ur Optik, Information und Photonik,
  Universit\"at Erlangen-N\"urnberg, 91058 Erlangen, Germany\\
\IEEEauthorrefmark{4}Max-Planck-Institut f\"ur die Physik des Lichts,
  Leuchs Division, 91058 Erlangen, Germany\\
\IEEEauthorrefmark{5} Canadian Institute for Advanced Research, Toronto, 
  Ontario, M5G 1Z8, Canada\\}
}


%


\maketitle


\maketitle

\begin{abstract}
We discuss a method to adapt the codeword stabilized (CWS) quantum
code framework to the problem of finding asymmetric quantum codes. We
focus on the corresponding Pauli error models for amplitude damping
noise and phase damping noise. In particular, we look at codes for
Pauli error models that correct one or two amplitude damping errors.
Applying local Clifford operations on graph states, we are able to
exhaustively search for all possible codes up to length $9$.  With a
similar method, we also look at codes for the Pauli error model that
detect a single amplitude error and detect multiple phase damping
errors.  Many new codes with good parameters are found, including
nonadditive codes and degenerate codes.
\end{abstract}


\section{Introduction}
\label{sec:intro}

Codeword stabilized (CWS) quantum codes constitute the by far most
general systematic framework for constructing quantum error-correcting
codes (QECC)~\cite{CSS+09,CZC08,CCS+09}. It encompasses stabilizer
codes~\cite{CRSS97,thesis:gottesman,CS96,steane96}, as well as many
nonadditive codes with good
parameters~\cite{RHSS97,yu2008nonadditive,lang2007nonadditive}. Over
the past years, it has been explored in various settings and has been
applied in many different cases, leading to promising
results~\cite{grassl-roetteler-2008a,grassl-roetteler-2008,looi-code,hu2008graphical,grassl2009generalized,li2010clustered,BCG+11,wang2013stabilizer}.

Most of the QECC constructed so far are for the depolarizing channel 
\begin{equation}
\mathcal{E}_{\text{DP}}(\rho)=(1-p)\rho +\frac{p}{3}(X\rho X+Y\rho Y +Z\rho Z),
\end{equation}
where the Pauli $X,Y,Z$ errors happen equally likely.  (Here $\rho$
denotes the density matrix representing the state of the quantum
system.)  The most general quantum channels allowed by quantum
mechanics are completely positive, trace-preserving linear maps that
can be represented in the Kraus decomposition
$\mathcal{E}(\rho)=\sum_k E_k\rho E_k^{\dag} $ with
$\sum_k E_k^{\dag}E_k=I$~\cite{nielsen2010quantum}.

One example generalizing the depolarizing channel $\mathcal{E}_{\text{DP}}$
is the asymmetric Pauli channel which sends $\rho$ to
\begin{equation}
\label{eq:DP}
(1-p_x-p_y-p_z)\rho +p_xX\rho X+p_yY\rho Y +p_z Z\rho Z, 
\end{equation}
where the Pauli $X,Y,Z$ errors happen with probabilities
$p_x,p_y,p_z$, respectively~\cite{PhysRevA.75.032345}. Other
asymmetric channels studied in the literature include the amplitude
damping channel~\cite{chuang1997bosonic}
\begin{equation}
\mathcal{E}_{\text{AD}}(\rho)=A_0\rho A_0^{\dag}+A_1\rho A_1^{\dag}, 
\end{equation}
where
\begin{equation}
\label{eq:AD}
A_0=\begin{pmatrix} 1 & 0 \\ 0 & \sqrt{1-\gamma} \end{pmatrix},
\quad
A_1=\begin{pmatrix} 0 & \sqrt{\gamma} \\ 0 & 0 \end{pmatrix},
\end{equation} 
for some damping parameter $\gamma$.

It has been demonstrated that designing QECC adaptively to specific
error models can result in better
codes~\cite{chuang1997bosonic,leung1997approximate,lang2007nonadditive,fletcher2008channel,sarvepalli2008asymmetric,shor2011high,duan2010multi,ezerman2013asymmetric}
and fault-tolerant protocols~\cite{aliferis2008fault}.  Although most
of these codes are indeed CWS codes, there has been no systematic
construction applying the CWS framework.  In this work, we fill this
gap by developing a method for finding CWS codes for asymmetric
channels.  Our method leads to many new codes with good parameters,
including nonadditive codes and degenerate codes. These results
demonstrate the power of the CWS framework for constructing good QECC.

\section{Error Models}
Depending on the noise model in different physical systems, we obtain
different asymmetric quantum channels.  We start with the amplitude
damping channel $\mathcal{E}_{\text{AD}}$ with Kraus operators given
in Eq.~\eqref{eq:AD}, which models real physical processes such as
spontaneous emission.  If the system is at finite temperature, then
the noise model will not only contain the Kraus operator $A_1$
corresponding to emission, but also $A_1^{\dag}$ corresponding to
absorption~\cite{nielsen2010quantum}. Notice that
\begin{equation}
\label{eq:A1}
A_1=\frac{\sqrt{\gamma}}{2}(X+iY),\quad
A_1^{\dag}=\frac{\sqrt{\gamma}}{2}(X-iY).
\end{equation}
Hence, the linear span of the operators $A_1$ and $A_1^{\dag}$
equals the linear span of $X$ and $Y$.  We can then equivalently
formulate the error model by using the Pauli operators $X$ and $Y$,
which happen with equal probability. That is, if a code is capable of
correcting $t$ $X$- and $t$ $Y$-errors, it can also correct $t$ $A_1$-
and $t$ $A_1^{\dag}$-errors.

Furthermore, notice that
\begin{eqnarray}
\label{eq:A0}
A_0=I-\frac{\gamma}{4}(I-Z)+O(\gamma^2),
\end{eqnarray}
while $A_1$ depends linearly on $\sqrt{\gamma}$. This then results in
an asymmetry between the probabilities $p_x=p_y$ and $p_z$ that
the Pauli $X,Y$ errors or the Pauli $Z$ error, respectively, happen.

Apart from amplitude damping, another common noise in physical systems
is dephasing, with Kraus operators given by $\sqrt{1-p}I$ and
$\sqrt{p}(I\pm Z)/2$, or equivalently, in terms of $I$ and $Z$ with
$p_z>0$ and $p_x=p_z=0$ \cite{nielsen2010quantum}.  In general, the
system undergoes both amplitude damping and dephasing, resulting in a
wide range for the parameters $p_x=p_y$ and $p_z$.

Therefore, in this work we consider the following asymmetric Pauli channel
\begin{alignat}{5}
\label{eq:AS}
\mathcal{E}_{\text{AS}}(\rho)={}&(1-(2p_{xy}+p_z))\rho\nonumber\\
&{}+p_{xy}(X\rho X+Y\rho Y)+p_z Z\rho Z,
\end{alignat}
where $X$ and $Y$ happen with equal probability $p_x=p_y=p_{xy}$. 

In terms of Eq.~\eqref{eq:AS}, the asymmetric Pauli error model
corresponding to amplitude damping is given by $p_{xy}\propto{\gamma}$
and $p_z\propto \gamma^2$. This is different from the amplitude
damping error model in, e.g.,
\cite{leung1997approximate,lang2007nonadditive,duan2010multi,shor2011high,GWYZ14},
where the Kraus operators $A_0$ and $A_1$ are used. The main reason
that we use the Pauli Kraus operators as our error sets is that this
enables us to use the CWS framework to construct codes.  Within the
CWS framework, in order to transform the quantum error detection
condition into a classical condition, it is more convenient to use
Pauli errors, as we will discuss in Sec.~\ref{sec:Alg}.  In other
words, since $A_0$ and $A_1$ are not Pauli operators, the CWS
framework does not directly apply.  Furthermore, due to
Eq.~\eqref{eq:A1} and Eq.~\eqref{eq:A0}, our error model does not only
correct the errors $A_0$ and $A_1$, but the resulting codes will be
stronger in the sense that $A^{\dag}_1$ can be corrected as well.

In this work we consider three specific cases for asymmetric codes, as
listed below.  We use $X_i,Y_i,Z_i$ to denote the Pauli $X,Y,Z$
operators on the $i$th qubit.  Notice that our method for generating
the error sets is very general and can be straightforwardly
generalized to deal with different relations between $p_x$, $p_y$, and
$p_z$.
\begin{enumerate}
\item[1.] Codes correcting a single amplitude damping error: to
  improve the fidelity of the transmitted state from $1-\gamma$ to
  $1-\gamma^2$, one only needs to correct a single $A_1$ error and
  detect a single $A_0$ error \cite{GWYZ14}.  In terms of Pauli
  operators, the corresponding error set is given by
\begin{equation}
\mathcal{E}^{\{1\}}=\{I\} \cup
\{X_i,Y_i,Z_i,\; X_iX_j,Y_iY_j,X_iY_j,Y_i Y_j\},
\end{equation}
where $i,j=1,\ldots,n$. A code that detects this error set in
fact also corrects a single $A_1^{\dag}$ error.
\item[2.] Codes correcting two amplitude damping errors: based
on the analysis on the single error case above, the error set is given by
\begin{equation}
\mathcal{E}^{\{2\}}=\{E_{\mu}E_{\nu}\colon E_{\mu},E_{\nu}\in\mathcal{E}^{\{1\}}\}.
\end{equation}
A code that detects this error set in fact also corrects two $A_1^{\dag}$ errors.
\item[3.] Codes detecting both a single amplitude damping error and
  multiple dephasing errors: detecting $\{X_i,Y_i,Z_i\}$ suffices to
  detect an arbitrary single qubit error (including a single amplitude
  damping error), and detecting all $Z$-errors up to weight $r$ will
  allow to correct $\lfloor r/2\rfloor$ $Z$-errors.  The error set is
\begin{equation}
\mathcal{E}^{\{3\}}=\{I\} \cup \{X_i,Y_i\colon i=1,\ldots,n\}\cup\mathcal{Z}_r,
\end{equation}
where $\mathcal{Z}_r$ is the set of all Pauli $Z$ operators up to
weight $r$.  A code that detects this error set in fact detects both an
arbitrary error and $r$ phase errors.
\end{enumerate}

\section{Algorithm to search for CWS codes}
\label{sec:Alg}

A QECC $Q$ is a subspace of the space of $n$ qubits
$(\mathbb{C}^2)^{\otimes n}$ (here we focus on quantum systems of
dimension $q=2$, but the approach can be generalized to qudits of
dimension $q>2$).  For a $K$-dimensional code space spanned by the
orthonormal basis $\{\ket{\psi_i}\colon i=1,\ldots,K\}$ and an error
set $\mathcal{E}$, there is a physical operation detecting all the
elements $E_{\mu}\in\mathcal{E}$ (as well as their linear
combinations) if the error detection
condition~\cite{bennett1996mixed,knill1997theory}
\begin{equation}\label{eq:codeConditions}
\bra{\psi_i}E_{\mu}\ket{\psi_j}=c_{\mu}\delta_{ij}, \quad c_\mu\in\C,
\end{equation} 
is satisfied.  The notation $((n,K))$ is used to denote a QECC with
length $n$ and dimension $K$.

Our goal is to find good codes detecting the error sets
$\mathcal{E}^{\{j\}}$, for each of the three cases.  For each code
length $n$, we seek the largest dimension $K$ of CWS codes for each
error set $\mathcal{E}^{\{j\}}$, $j=1,2,3$. This is done through a
maximum clique search~\cite{CCS+09}, by using the algorithms and
programs developed in~\cite{xiang2013scalable}.

\subsection{The CWS framework}

An $((n,K))$ CWS code $Q$ is described by two objects: 1) A stabilizer
$S$ that is an abelian subgroup of the $n$-qubit Pauli group, has
order $2^n$, and does not contain $-I$; the group $S$ is called the
word stabilizer. 2) A set of $K$ $n$-qubit Pauli operators
$W=\{w_\ell\colon \ell=1,\ldots,K\}$, which are called the word
operators.  There is a unique quantum state $\ket{S}$ stabilized by
$S$, i.e., $s\ket{S}=\ket{S}$ for all $s\in S$. The code $Q$ is then
spanned by the basis vectors given by $\ket{w_\ell}=w_\ell\ket{S}$.

According to Eq.~\eqref{eq:codeConditions}, the code $Q$ detects the
error set $\mathcal{E}$ if and only if
$\bra{w_i}E\ket{w_j}=c_{E}\delta_{ij}$ for all $E\in\mathcal{E}$. When
$\mathcal{E}$ consists of Pauli matrices, this error-detecting
condition can be written in terms of $S$ and $w_i$ as
below~\cite{CSS+09}:

For all $E\in\mathcal{E}$,
\begin{eqnarray}
\forall i\neq j\colon w_i^{\dag}E w_j\notin\pm S
\end{eqnarray}
and
\begin{alignat}{5}
\label{eq:nondeg}
&\left(\forall i\colon w_i^{\dag}E w_i\notin\pm S\right)&\quad\text{or}\\
&\left(\forall i\colon w_i^{\dag}E w_i\in S\right)&\quad\text{or}\\
&\left(\forall i\colon w_i^{\dag}E w_i\in -S\right).
\end{alignat}
If condition \eqref{eq:nondeg} holds for all $E\in\mathcal{E}$
different from identity, then the code $Q$ is nondegenerate, otherwise
it is degenerate.

\subsection{The CWS standard form}

Every $((n,K))$ CWS code can be transformed, by local Clifford operations,
into a standard form~\cite{CSS+09}, where the word operators take the form
$
w_\ell=Z^{\mathbf{c}_\ell}
$
and the word stabilizer has generators of the form
$
S_i=X_iZ^{\mathbf{r}_i},
$
for some choices of classical $n$-bit strings ${\mathbf{c}_{\ell}}$ and
${\mathbf{r}_i}$. Here
$Z^{\mathbf{c}_\ell}=Z^{c_{\ell,1}}\otimes\ldots\otimes Z^{c_{\ell,n}}$.

In the standard form, any $n$-qubit Pauli error, which can be written
in the form $E=\pm Z^{\mathbf{v}}X^{\mathbf{u}}$ for some classical
$n$-bit strings $\mathbf{v}$ and $\mathbf{u}$, can be translated to
classical errors via the map
\begin{equation}
\text{Cl}_S(E=\pm Z^{\mathbf{v}}X^{\mathbf{u}})=\mathbf{v}\oplus\bigoplus_{i=1}^n (\mathbf{u})_i\mathbf{r}_i.
\end{equation}

Now for the word operators
$\{Z^{\mathbf{c}_\ell}\colon{\mathbf{c}_\ell\in\mathcal{C}}\}$, the
error detection condition requires that the classical binary code
$\mathcal{C}$ detects all errors from $\text{Cl}_S(\mathcal{E})$, and
that for each $E\in\mathcal{E}$
\begin{alignat}{5}\label{eq:deg2}
&&\text{Cl}_S(E)&{}\neq\mathbf{0}\\
\text{or}\quad&&\forall\ell\colon Z^{\mathbf{c}_\ell} E&{}=EZ^{\mathbf{c}_\ell}.
\end{alignat}
If Eq.~\eqref{eq:deg2} holds for all $E\in\mathcal{E}$, the CWS code
is nondegenerate, otherwise it is degenerate.

\subsection{Local Clifford operations}
To get to the standard form, one needs to apply local Clifford (LC)
operations of the form $L=\bigotimes_i L_i$, where $L_i$ are single-qubit
Clifford operations~\cite{CSS+09}.  This transforms the stabilizer $S$
and word operators $\{w_\ell\}$ to the standard form, but at the same
time also changes the error model.

For the depolarizing channel given in Eq.~\eqref{eq:DP}, the error set
is invariant under LC operations, since in this model essentially all
single-qubit errors happen equally likely. Therefore, in order to
search for a CWS code, one can simply use the standard form by
starting from a stabilizer of the form $S_i=X_iZ^{\mathbf{r}_i}$, which corresponds to a graph
state~\cite{PhysRevA.69.062311}. For a fixed length $n$, it is
sufficient to consider all graph states up to LC equivalence as
classified in~\cite{danielsen2006classification}.  This results in an
exhaustive search for all possible CWS codes of length $n$.

Being able to restrict the search to graph states up to LC
equivalence, instead of all stabilizer states of length $n$, has
dramatically reduced the search space, and exhaustive search for
single-error-correcting codes for the depolarizing channel up to
length $n=10$ has been carried out. It turned out that the best CWS
code with length $n=9$ has dimension $K=12$, beating the best
stabilizer code of dimension $2^3=8$~\cite{yu2008nonadditive}; for
$n=10$ the best CWS code has dimension $K=24$, again beating the best
stabilizer code of dimension $2^4=16$~\cite{hu2008graphical}.

For the asymmetric channels as given in Eq.~\eqref{eq:AS}, however,
considering only all graphs states as classified
in~\cite{danielsen2006classification} and the error sets
$\mathcal{E}^{\{j\}}$ ($j=1,2,3$) is not sufficient to exhaustively
search for all possible CWS codes.  This is due to the asymmetry
between $p_{xy}$ and $p_z$, which implies that the error sets are no
longer invariant under LC operations.  Therefore, in order to
exhaustively search for all possible CWS codes by using the standard
form, one will need to check all the possible error sets that are LC
equivalent to a given $\mathcal{E}^{\{j\}}$.

Recall that the single-qubit Clifford group is generated by the
Hadamard operator $H$ and the phase operator $P$ as given
below~\cite{CRSS97,thesis:gottesman}
\begin{equation}
H=\frac{1}{\sqrt{2}}\begin{pmatrix}1 & 1 \\ 1 & -1\end{pmatrix},\quad
P=\begin{pmatrix}1 & 0 \\ 0 & i\end{pmatrix}.
\end{equation}
Since overall phase factors can be ignored, we only need to consider
the action of the Clifford group on the Pauli matrices $X,Y,Z$ modulo
phase factors.  The Clifford group acts as the permutation group $S_3$
on three letters (we use $1,2,3$ to denote $X,Y,Z$, respectively). The
group $S_3$ has order six, with the elements given by (in cycle
notation) $\{\text{id},(123),(132),(12),(13),(23)\}$, where
$\text{id}$ denotes the identity permutation. All error sets
$\mathcal{E}^{\{j\}}$, $j=1,2,3$, are invariant with respect to
interchanging $X$ and $Y$. Hence it is sufficient to consider one
representative from each of the three right cosets of $(12)$, given by
$\{\text{id},(12)\}$, $\{(13),(132)\}$, and $\{(23),(123)\}$.  So
effectively, we only need to test, e.g., the three permutations
$\{\text{id},(13),(23)\}$.

Therefore, for each of the error sets $\mathcal{E}^{\{j\}}$, we have
three cases for each qubit $i$: no permutation, permute $Y$ and $Z$,
or permute $X$ and $Z$.  To search for a length $n$ code, this will
reduce the total number of error sets from $6^n$ to $3^n$ for each
graph state to be tested.  Compared to codes for the depolarizing
channel, the search space is enlarged by a factor of $3^n$, due to the
asymmetry between $p_{xy}$ and $p_z$.  Nevertheless, we can still
handle the search for small $n$, in particular for the error sets
$\mathcal{E}^{\{j\}}$, $j=1,2,3$, up to length $n=9$.

\section{Results}
As described above, in our search algorithm, we start from the CWS
standard form and transform the error set $\mathcal{E}^{\{j\}}$ by LC
operations.  This has no effect on the code parameters $((n,K))$
found.  However, to present the CWS codes found, we fix the error set
$\mathcal{E}^{\{j\}}$ and equivalently transform the CWS standard form
into a general CWS code.

\subsection{Codes correcting a single amplitude damping error}

We have conducted an exhaustive search for the error set
$\mathcal{E}^{\{1\}}$ up to length $n=9$, resulting in CWS codes
correcting a single amplitude damping error.  As already mentioned,
the codes found can not only correct a single error given by the Kraus
operator $A_1$, at the same time they also correct a single error
$A_1^{\dag}$.  In other words, the codes correct both single $X$
errors and single $Y$ errors (and detect single $Z$ errors as well).
We summarize our results in Table~\ref{tb:1AD}.

\begin{table}[hbt]
\def\mycaption{\parbox{\hsize}{\normalfont\footnotesize
  Dimension $K$ of CWS codes $((n,K))$ of length
  $n$ detecting the error set $\mathcal{E}^{\{1\}}$ for different
  length $n$.  The column $d=3$ lists the largest dimension of CWS
  codes that correct a single error for the depolarizing channel.  The
  column $\mathcal{E}^{\{1\}}$ lists the largest dimension of CWS codes
  found detecting the error set $\mathcal{E}^{\{1\}}$. The column CSS
  lists the largest dimension of the known Calderbank-Shor-Steane (CSS)
  codes~\cite{CRSS97,steane96} that can correct the error set
  $\mathcal{E}^{\{1\}}$, based on a construction proposed
  in~\cite{thesis:gottesman}. The column GF(3) lists the largest
  dimension of codes correcting a single amplitude damping error based
  on a construction proposed in~\cite{shor2011high}.}}
\caption{\protect\smallskip\protect\mycaption}\label{tb:1AD}
\centerline{%
  \begin{tabular}{| c || c | c | c | c |}
    \hline
    $n$ & $d=3$~\cite{CRSS97,yu2008nonadditive} & $\mathcal{E}^{\{1\}}$ & CSS~\cite{thesis:gottesman} & GF(3)~\cite{shor2011high}\\ \hline
      &   &   &   &   \\[-2ex]\hline
    5 & 2 & 2 & 2 & 2 \\ \hline
    6 & 2 & 4 & 2 & 5 \\\hline
    7 & 2 & 8 & 8 & 8 \\\hline
    8 & 8 & 10 & 8 & 16 \\\hline
    9 & 12 & 20 & 16 & 24 \\
    \hline
  \end{tabular}}
\end{table}

As we can see from the table, for the lengths $6,7,8,9$, our codes
outperform the best single-error-correcting codes for the depolarizing
channel---which also correct the error set $\mathcal{E}^{\{1\}}$,
i.e., a single amplitude damping error.  In particular, for lengths $8$
and $9$, the best CWS codes we have found (of dimensions $10$ and $20$
respectively) are nonadditive codes.

For lengths $6,8,9$, our codes also outperform the best known CSS
codes that are specifically designed to detect the error set
$\mathcal{E}^{\{1\}}$, based on a construction proposed
in~\cite{thesis:gottesman} (see also~\cite{shor2011high}). Therefore,
for these lengths, we have found good codes that outperform all the
previously known constructions for detecting the error set
$\mathcal{E}^{\{1\}}$.

Notice that the existence of a CWS code with dimension $K=4$, and
hence a subcode of dimension $3$, implies the existence of a
stabilizer code with the same parameters~\cite[Theorem 7]{CCS+09}.
Hence the $((6,4))$ codes we found, as listed in Table~\ref{tb:1AD},
include stabilizer codes encoding two qubits.  As an example, one such
code has stabilizer $S$ generated by
\begin{center}\tabcolsep0.5\tabcolsep
\begin{tabular}{cccccc}
$X$ & $X$ & $I$ & $I$ & $Z$ & $Z$ \\
$X$ & $Z$ & $I$ & $Z$ & $I$ & $X$ \\
$Z$ & $I$ & $Y$ & $Z$ & $Y$ & $Z$ \\
$I$ & $I$ & $Z$ & $X$ & $I$ & $Z$ 
\end{tabular}
\end{center}
It is straightforward to check that this code detects the error set
$\mathcal{E}^{\{1\}}$,
since no elements in $\mathcal{E}^{\{1\}}$ is also in $C(S)\setminus
S$, where $C(S)$ is the centralizer of the stabilizer $S$.

However, with the exception of $n=7$, the single-error-correcting
codes constructed in~\cite{shor2011high} have larger dimensions than
our codes.  The codes constructed in~\cite{shor2011high} are
specifically designed to correct the Kraus operators $A_0$ and $A_1$,
these codes cannot detect the error set $\mathcal{E}^{\{1\}}$.  As
detection of the errors $\mathcal{E}^{\{1\}}$ implies that a single
error $A_1^{\dag}$ can be corrected as well, it is not a surprise that
our codes have smaller dimensions.

Notice that the codes constructed in~\cite{shor2011high} are also CWS
codes, but errors are handled in a different way than the Pauli error
set $\mathcal{E}^{\{1\}}$.  It remains open how to generalize the
method of \cite{shor2011high} to deal with more than one amplitude
damping error, while the error set $\mathcal{E}^{\{1\}}$ can naturally
be generalized, e.g., to $\mathcal{E}^{\{2\}}$ for correcting two
amplitude damping errors, as demonstrated next.

\subsection{Codes correcting two amplitude damping errors}

We have performed an exhaustive search for codes correcting two
amplitude damping errors, i.e., detecting the error set
$\mathcal{E}^{\{2\}}$, up to length $n=9$. In fact, the resulting
codes correct any combination of $X$ and $Z$ errors up to weight two,
as well as a single $Z$ error.

No non-trivial CWS codes are found for length $n\leq 8$.  For length
$n=9$, two LC-inequivalent codes encoding a single qubit have been
found found.  These are both stabilizer codes encoding a single qubit,
since the corresponding classical code $\mathcal{C}$ is trivially
linear~\cite{CSS+09}.

One code has the stabilizer $S_1$ generated by
\begin{center}\tabcolsep0.5\tabcolsep
\begin{tabular}{lllllllll}
$X$ &$I$ &$I$ &$I$ &$I$ &$I$ &$I$ &$I$ &$Z$\\
$Z$ &$I$ &$I$ &$I$ &$X$ &$Z$ &$I$ &$Z$ &$X$\\
$I$ &$X$ &$I$ &$I$ &$I$ &$I$ &$I$ &$Z$ &$I$\\
$I$ &$Z$ &$I$ &$I$ &$X$ &$I$ &$Z$ &$X$ &$Z$\\
$I$ &$I$ &$X$ &$I$ &$I$ &$Z$ &$I$ &$I$ &$I$\\
$I$ &$I$ &$Z$ &$I$ &$Y$ &$Y$ &$Z$ &$Z$ &$I$\\
$I$ &$I$ &$I$ &$X$ &$I$ &$I$ &$Z$ &$I$ &$I$\\
$I$ &$I$ &$I$ &$Z$ &$Y$ &$Z$ &$Y$ &$I$ &$Z$ 
\end{tabular}
\end{center}

The other code has the stabilizer $S_2$ generated by
\begin{center}\tabcolsep0.5\tabcolsep
\begin{tabular}{lllllllll}
$X$ &$I$ &$I$ &$I$ &$I$ &$I$ &$I$ &$I$ &$Z$\\
$Z$ &$I$ &$I$ &$I$ &$Z$ &$Z$ &$Z$ &$I$ &$X$\\
$I$ &$X$ &$I$ &$I$ &$I$ &$I$ &$I$ &$Z$ &$I$\\
$I$ &$Z$ &$I$ &$Z$ &$Z$ &$Y$ &$I$ &$Y$ &$Z$\\
$I$ &$I$ &$X$ &$I$ &$I$ &$I$ &$Z$ &$I$ &$I$\\
$I$ &$I$ &$Z$ &$Z$ &$Z$ &$X$ &$X$ &$Z$ &$I$\\
$I$ &$I$ &$I$ &$X$ &$I$ &$Z$ &$I$ &$I$ &$I$\\
$I$ &$I$ &$I$ &$I$ &$X$ &$I$ &$Z$ &$Z$ &$Z$ 
\end{tabular}
\end{center}

It is straightforward to check that these codes detect the error set
$\mathcal{E}^{\{2\}}$, since no elements in $\mathcal{E}^{\{2\}}$ is
also in $C(S_i)\setminus S_i$ (for $i=1,2$). Furthermore, both codes
are degenerate since some of the elements in $\mathcal{E}^{\{2\}}$ are
indeed in $S_i$, for instance $X_1Z_9$.

These codes outperform the $((10,2))$ code found
in~\cite{duan2010multi}. Recall that Shor's nine-qubit code, having
the same parameters $((9,2))$ as our codes, also corrects two amplitude
damping errors \cite{thesis:gottesman}. However, Shor's code only
corrects the Kraus operators $A_0$ and $A_1$, but does not detect the
error set $\mathcal{E}^{\{2\}}$.  Therefore, for length $n=9$, we have
found good codes that outperform all the previous known constructions
for detecting the error set $\mathcal{E}^{\{2\}}$.

\subsection{Codes detecting a single amplitude damping error and detecting multiple dephasing errors}

For the error set $\mathcal{E}^{\{3\}}$, we have performed an
exhaustive search for different lengths $n$ and $Z$-error detecting
capabilities $r$ up to $n=8$, and a random search starting from
randomly selected graph states for $n=9$ and different $r$.  Our
results are listed in Table~\ref{tb:de1AD}. We compare our results
with the best stabilizer codes that detect all errors up to weight
$r$ as given in~\cite{CRSS97}, and the codes detecting a single
amplitude damping errors and $Z$ errors up to weight $r$ as found
in~\cite{ezerman2013asymmetric}.

\begin{table*}[hbt]
\def\mycaption{\parbox{\hsize}{\normalfont\footnotesize\rule{0pt}{2.5ex}%
    Dimension $K$ of CWS codes detecting the error set
    $\mathcal{E}^{\{3\}}$ for different length $n$ and parameter
    $r$. For each value of $r$, the first column lists the largest
    dimension of stabilizer codes that detect all errors up to weight
    $r$ as given in~\cite{CRSS97}; the second column lists the largest
    dimension of asymmetric codes detecting a single amplitude damping
    error and phase errors up to weight $r$ as found
    in~\cite{ezerman2013asymmetric}; the third column lists the
    largest dimension of the CWS codes found by our search for codes
    detecting the error set $\mathcal{E}^{\{3\}}$.  `$-$' means that
    no non-trivial codes exist based on the construction. The numbers
    labeled with $^*$ are the best parameters found by random search;
    otherwise the maximal dimension is obtained by exhaustive search. }}
\caption{\protect\mycaption}
\begin{center}\def\arraystretch{1.1}\def\txt#1{\text{\scriptsize\!\! #1\!\!}}
  $\begin{array}{| c || c|c|c || c|c|c || c|c|c || c|c|c || c|c|c || c|c|c || c|c|c |}
    \hline
    n/r & \multicolumn{3}{c||}{1}& \multicolumn{3}{c||}{2}& \multicolumn{3}{c||}{3}& \multicolumn{3}{c||}{4}& \multicolumn{3}{c||}{5}& \multicolumn{3}{c||}{6}& \multicolumn{3}{c|}{7}\\ \hline
&\multicolumn{3}{c||}{}&\multicolumn{3}{c||}{}&\multicolumn{3}{c||}{}&\multicolumn{3}{c||}{}&\multicolumn{3}{c||}{}&\multicolumn{3}{c||}{}&\multicolumn{3}{c|}{}\\[-2.5ex]
    \hline
& \txt{stab.} &\txt{\cite{ezerman2013asymmetric}}&\txt{CWS}&\txt{stab.} &\txt{\cite{ezerman2013asymmetric}}&\txt{CWS}&\txt{stab.} &\txt{\cite{ezerman2013asymmetric}}&\txt{CWS}&\txt{stab.} &\txt{\cite{ezerman2013asymmetric}}&\txt{CWS}&\txt{stab.} &\txt{\cite{ezerman2013asymmetric}}&\txt{CWS}&\txt{stab.} &\txt{\cite{ezerman2013asymmetric}}&\txt{CWS}&\txt{stab.} &\txt{\cite{ezerman2013asymmetric}}&\txt{CWS}\\\hline
    5 & 4& 5 & 6 & 2& - & 4& - & - & 2& - & - & 2& - & - & - & - & - & - & - & - & - \\\hline
    6 & 16 & 16 & 16 & 2& 2& 8 & - & - & 4& - & - & 2& - & - & 2& - & - & - & - & - & - \\\hline
    7 & 16 & 22& 24 & 2& 8 & 16 & - & - & 8 & - & - & 2& - & - & 2& - & - & 2& - & - & - \\\hline
    8 & 64 & 64 & 64 & 8 & 8 & 20 & - & 8 & 16 & - & - & 4& - & - & 2& - & - & 2& - & - & 2\\\hline
    9 & 64 & 93 & 96\rlap{${}^*$}& 8 & 16 & 40\rlap{${}^*$} & - & 8 & 20\rlap{${}^*$} & - & - & 6\rlap{${}^*$} & - & - & 4\rlap{${}^*$} & - & - & 2\rlap{${}^*$} & - & - & 2\rlap{${}^*$} \\\hline        
  \end{array}$
\end{center}  
\label{tb:de1AD}
\end{table*}

As we can see from the table, for most lengths $n$ and $Z$-error
weight $r$, the CWS codes found outperform the known results. The
entries for which we did not find improvements are $n=6$, $r=1$ and
$n=8$, $r=1$. Codes with $r=1$ detect single Pauli errors, i.e., they
are codes of minimum distance two.  For even length, the corresponding
stabilizer codes are known to have the largest possible dimension for
single-error-detecting codes~\cite{rains1999quantum}.  For odd length
$n=5,7,9$, we find codes with parameters matching those of the code
family $((2m+1,3\times 2^{2m-3},2))$ given
in~\cite{rains1999quantum}. Whenever the dimension is a power of two,
the codes we found include stabilizer codes.

These results demonstrate the power of the CWS framework for
constructing good QECC, even with random search.

\section*{Acknowledgements} We thank Dr. Jingen Xiang
for providing us computer programs
for maximum clique search. TJ and BZ are supported by NSERC.



%

\bibliographystyle{IEEEtranS}
\bibliography{CWSAsy}

\begin{thebibliography}{10}
\providecommand{\url}[1]{#1}
\csname url@samestyle\endcsname
\providecommand{\newblock}{\relax}
\providecommand{\bibinfo}[2]{#2}
\providecommand{\BIBentrySTDinterwordspacing}{\spaceskip=0pt\relax}
\providecommand{\BIBentryALTinterwordstretchfactor}{4}
\providecommand{\BIBentryALTinterwordspacing}{\spaceskip=\fontdimen2\font plus
\BIBentryALTinterwordstretchfactor\fontdimen3\font minus
  \fontdimen4\font\relax}
\providecommand{\BIBforeignlanguage}[2]{{%
\expandafter\ifx\csname l@#1\endcsname\relax
\typeout{** WARNING: IEEEtranS.bst: No hyphenation pattern has been}%
\typeout{** loaded for the language `#1'. Using the pattern for}%
\typeout{** the default language instead.}%
\else
\language=\csname l@#1\endcsname
\fi
#2}}
\providecommand{\BIBdecl}{\relax}
\BIBdecl

\bibitem{aliferis2008fault}
P.~Aliferis and J.~Preskill, ``Fault-tolerant quantum computation against
  biased noise,'' \emph{Physical Review A}, vol.~78, no.~5, p. 052331, 2008.

\bibitem{BCG+11}
S.~{Beigi}, I.~{Chuang}, M.~{Grassl}, P.~{Shor}, and B.~{Zeng}, ``{Graph
  concatenation for quantum codes},'' \emph{Journal of Mathematical Physics},
  vol.~52, no.~2, p. 022201, 2011.

\bibitem{bennett1996mixed}
C.~H. Bennett, D.~P. DiVincenzo, J.~A. Smolin, and W.~K. Wootters,
  ``Mixed-state entanglement and quantum error correction,'' \emph{Physical
  Review A}, vol.~54, no.~5, pp. 3824--3851, 1996.

\bibitem{CRSS97}
A.~R. Calderbank, E.~Rains, P.~W. Shor, and N.~J.~A. Sloane, ``Quantum error
  correction via codes over ${GF}(4)$,'' \emph{IEEE Transactions on Information
  Theory}, vol.~44, no.~4, pp. 1369--1387, 1998.

\bibitem{CS96}
A.~R. {Calderbank} and P.~W. {Shor}, ``{Good quantum error-correcting codes
  exist},'' \emph{Physcial Review A}, vol.~54, no.~2, pp. 1098--1105, 1996.

\bibitem{CZC08}
X.~{Chen}, B.~{Zeng}, and I.~L. {Chuang}, ``{Nonbinary codeword-stabilized
  quantum codes},'' \emph{Physical Review A}, vol.~78, no.~6, p. 062315, 2008.

\bibitem{CCS+09}
I.~{Chuang}, A.~{Cross}, G.~{Smith}, J.~{Smolin}, and B.~{Zeng}, ``{Codeword
  stabilized quantum codes: Algorithm and structure},'' \emph{Journal of
  Mathematical Physics}, vol.~50, no.~4, p. 042109, 2009.

\bibitem{chuang1997bosonic}
I.~L. Chuang, D.~W. Leung, and Y.~Yamamoto, ``Bosonic quantum codes for
  amplitude damping,'' \emph{Physical Review A}, vol.~56, no.~2, pp.
  1114--1125, 1997.

\bibitem{CSS+09}
A.~Cross, G.~Smith, J.~A. Smolin, and B.~Zeng, ``Codeword stabilized quantum
  codes,'' \emph{IEEE Transactions on Information Theory}, vol.~55, no.~1, pp.
  433--438, 2009.

\bibitem{danielsen2006classification}
L.~E. Danielsen and M.~G. Parker, ``On the classification of all self-dual
  additive codes over {GF(4)} of length up to 12,'' \emph{Journal of
  Combinatorial Theory, Series A}, vol. 113, no.~7, pp. 1351--1367, 2006.

\bibitem{duan2010multi}
R.~Duan, M.~Grassl, Z.~Ji, and B.~Zeng, ``Multi-error-correcting amplitude
  damping codes,'' in \emph{Proceedings 2010 IEEE International Symposium on
  Information Theory (ISIT 2010)}, 2010, pp. 2672--2676.

\bibitem{ezerman2013asymmetric}
M.~F. Ezerman and M.~Grassl, ``Asymmetric quantum codes detecting a single
  amplitude error,'' in \emph{Proceedings 2013 IEEE International Symposium on
  Information Theory (ISIT 2013)}, 2013, pp. 922--926.

\bibitem{fletcher2008channel}
A.~S. Fletcher, P.~W. Shor, and M.~Z. Win, ``Channel-adapted quantum error
  correction for the amplitude damping channel,'' \emph{IEEE Transactions on
  Information Theory}, vol.~54, no.~12, pp. 5705--5718, 2008.

\bibitem{thesis:gottesman}
D.~Gottesman, ``Stabilizer codes and quantum error correction,'' Ph.D.
  dissertation, California Institute of Technology, Pasadena, CA, 1997.

\bibitem{grassl-roetteler-2008a}
M.~Grassl and M.~R{\"o}tteler, ``Non-additive quantum codes from {G}oethals and
  {P}reparata codes,'' in \emph{Proceedings 2008 IEEE Information Theory
  Workshop (ITW 2008)}, 2008, pp. 396--400.

\bibitem{grassl-roetteler-2008}
------, ``Quantum {G}oethals-{P}reparata codes,'' in \emph{Proceedings 2008
  IEEE International Symposium on Information Theory (ISIT 2008)}, 2008, pp.
  300--304.

\bibitem{grassl2009generalized}
M.~Grassl, P.~Shor, G.~Smith, J.~Smolin, and B.~Zeng, ``Generalized
  concatenated quantum codes,'' \emph{Physical Review A}, vol.~79, no.~5, p.
  050306, 2009.

\bibitem{GWYZ14}
M.~Grassl, Z.~Wei, Z.-Q. Yin, and B.~Zeng, ``Quantum error-correcting codes for
  amplitude damping,'' in \emph{Proceedings 2014 IEEE International Symposium
  on Information Theory (ISIT 2014)}, 2014, pp. 906--910.

\bibitem{PhysRevA.69.062311}
M.~Hein, J.~Eisert, and H.~J. Briegel, ``Multiparty entanglement in graph
  states,'' \emph{Physical Review A}, vol.~69, p. 062311, 2004.

\bibitem{hu2008graphical}
D.~Hu, W.~Tang, M.~Zhao, Q.~Chen, S.~Yu, and C.~H. Oh, ``Graphical nonbinary
  quantum error-correcting codes,'' \emph{Physical Review A}, vol.~78, no.~1,
  p. 012306, 2008.

\bibitem{PhysRevA.75.032345}
L.~Ioffe and M.~M{\'e}zard, ``Asymmetric quantum error-correcting codes,''
  \emph{Physical Review A}, vol.~75, no.~3, p. 032345, 2007.

\bibitem{knill1997theory}
E.~Knill and R.~Laflamme, ``Theory of quantum error-correcting codes,''
  \emph{Physical Review A}, vol.~55, no.~2, pp. 900--911, 1997.

\bibitem{lang2007nonadditive}
R.~Lang and P.~W. Shor, ``Nonadditive quantum error correcting codes adapted to
  the amplitude damping channel,'' \emph{arXiv preprint arXiv:0712.2586}, 2007.

\bibitem{leung1997approximate}
D.~W. Leung, M.~A. Nielsen, I.~L. Chuang, and Y.~Yamamoto, ``Approximate
  quantum error correction can lead to better codes,'' \emph{Physical Review
  A}, vol.~56, no.~4, pp. 2567--2573, 1997.

\bibitem{li2010clustered}
Y.~Li, I.~Dumer, and L.~P. Pryadko, ``Clustered error correction of
  codeword-stabilized quantum codes,'' \emph{Physical Review Letters}, vol.
  104, no.~19, p. 190501, 2010.

\bibitem{looi-code}
S.~Y. Looi, L.~Yu, V.~Gheorghiu, and R.~B. Griffiths, ``Quantum error
  correcting codes using qudit graph states,'' \emph{Physical Review A},
  vol.~78, no.~4, p. 042303, 2008.

\bibitem{nielsen2010quantum}
M.~A. Nielsen and I.~L. Chuang, \emph{Quantum computation and quantum
  information}.\hskip 1em plus 0.5em minus 0.4em\relax Cambridge University
  Press, 2010.

\bibitem{RHSS97}
E.~M. Rains, R.~H. Hardin, P.~W. Shor, and N.~J.~A. Sloane, ``A nonadditive
  quantum code,'' \emph{Physical Review Letters}, vol.~79, no.~5, pp. 953--954,
  1997.

\bibitem{rains1999quantum}
E.~M. Rains, ``Quantum codes of minimum distance two,'' \emph{IEEE Transactions
  on Information theory}, vol.~45, no.~1, pp. 266--271, 1999.

\bibitem{sarvepalli2008asymmetric}
P.~K. Sarvepalli, A.~Klappenecker, and M.~R{\"o}tteler, ``Asymmetric quantum
  ldpc codes,'' in \emph{Proceedings 2008 IEEE International Symposium on
  Information Theory (ISIT 2008)}, 2008, pp. 305--309.

\bibitem{shor2011high}
P.~W. Shor, G.~Smith, J.~Smolin, and B.~Zeng, ``High performance
  single-error-correcting quantum codes for amplitude damping,'' \emph{IEEE
  Transactions on Information Theory}, vol.~57, no.~10, pp. 7180--7188, 2011.

\bibitem{steane96}
A.~Steane, ``Multiple particle interference and quantum error correction,''
  \emph{Proceedings of the Royal Society of London, Series A}, vol. 452, no.
  1954, pp. 2551--2577, 1996.

\bibitem{wang2013stabilizer}
Y.-J. Wang, B.~Zeng, M.~Grassl, and B.~C. Sanders, ``Stabilizer formalism for
  generalized concatenated quantum codes,'' in \emph{Proceedings 2013 IEEE
  International Symposium on Information Theory (ISIT 2013)}, 2013, pp.
  529--533.

\bibitem{xiang2013scalable}
J.~Xiang, C.~Guo, and A.~Aboulnaga, ``Scalable maximum clique computation using
  {MapReduce},'' in \emph{Proceedings 2013 IEEE 29th International Conference
  on Data Engineering (ICDE)}, 2013, pp. 74--85.

\bibitem{yu2008nonadditive}
S.~Yu, Q.~Chen, C.~Lai, and C.~Oh, ``Nonadditive quantum error-correcting
  code,'' \emph{Physical Review Letters}, vol. 101, no.~9, p. 090501, 2008.

\end{thebibliography}

\end{document}